\documentclass[a4paper,aps,prd,twocolumn,showkeys,showpacs,nofootinbib]{revtex4}
\usepackage{latexsym}
\usepackage{mathrsfs}
\usepackage{hyperref}
\usepackage{color}
\usepackage{csquotes}
\usepackage{amsbsy,amsmath,amssymb,calc,amsfonts}
\usepackage{graphicx,calc,epsfig}
\usepackage[english]{babel} 
\usepackage{mathtools}
\usepackage{bbm, bbold}
\usepackage{dsfont}
\hypersetup{
  colorlinks   = true, 
  urlcolor     = blue, 
  linkcolor    = blue, 
  citecolor   =  red 
}
\usepackage[multiple]{footmisc}

\usepackage{titlesec}
\titleformat{\paragraph}{\normalfont\normalsize}{\theparagraph}{1em}{}
\titlespacing*{\paragraph}
{0pt}{3.25ex plus 1ex minus .2ex}{1.5ex plus .2ex}

\def\ut#1{\rlap{\lower1ex\hbox{$\sim$}}#1{}}
\newcommand{\N}{\mathbb{N}}

\newcommand{\R}{\mathbb{R}}
\newcommand{\be}{\begin{equation}}
\newcommand{\ee}{\end{equation}}
\newcommand{\ba}{\nopagebreak[3]\begin{eqnarray}}
\newcommand{\ea}{\end{eqnarray}}
\renewcommand{\eqref}[1]{Eq.\,(\ref{#1})}

\newcommand{\std}{d}			
\newcommand{\sd}{D}				
\newcommand{\rk}{n}             
\newcommand{\cl}{\xi}	        
\newcommand{\gt}{C}             
\newcommand{\gth}{\hat{C}}      
\newcommand{\Sk}{S_0}           
\newcommand{\Sia}{S_\textsc{ia}}   
\newcommand{\gf}{\varphi}       

\newcommand{\SU}{\mathrm{SU}}

\newcommand{\SL}{\mathrm{SL}}

\newcommand{\tr}{\mathrm{tr}}
\newcommand{\id}{\mathbb 1}
\def\d{\mathrm{d}}

\begin{document}

\begin{flushleft}
KCL-PH-TH/2018-39
\end{flushleft}

\title{Phase transitions in group field theory:\\ The Landau perspective}

\date{\today}

\author{Andreas G. A. Pithis$^1$}\email{andreas.pithis@kcl.ac.uk}
\author{Johannes Th\"urigen$^{2}$}\email{johannes.thuerigen@physik.hu-berlin.de}
\affiliation{$^1$Theoretical Particle Physics and Cosmology Group, Department of Physics, King's College London, University of London, Strand, London, WC2R 2LS, U.K., EU\\ 
$^2$Institut~f\"ur~Physik,~Insitut f\"ur~Mathematik,~Humboldt-Universit\"at~zu~Berlin,~12489~Berlin,~Germany,~EU}

\thanks{}

\begin{abstract}
In various approaches to quantum gravity continuum spacetime is expected to emerge from discrete geometries through a phase transition.
In group field theory, various indications for such a transition have recently been found but a complete understanding of such a phenomenon remains an open issue.
In this work, 
we investigate the critical behavior of different group field theory models in the Gaussian approximation. Applying the Ginzburg criterion to quantify field fluctuations, 
we find that this approximation breaks down in the case of three-dimensional Euclidean quantum gravity as described by the dynamical Boulatov model on the compact group $\textrm{SU}(2)$. This result is independent of the peculiar gauge symmetry and specific form of nonlocality of the model.
On the contrary, we find that the Gaussian approximation is valid for a rank-$1$ GFT on the noncompact sector of fields on $\textrm{SL}(2,\mathbb{R})$ related to Lorentzian models.
Though a nonperturbative analysis is needed to settle the question of phase transitions for compact groups, the results may also indicate the necessity to consider group field theory on noncompact domains for phase transitions to occur.
\end{abstract}


\maketitle

\section{Introduction}

One of the hardest problems in approaches to quantum gravity based on discrete quantum geometries
is to recover continuum space and spacetime geometry, their symmetries and dynamics 
in an appropriate limit. 

A promising candidate for such a quantum theory of gravity is group field theory (GFT)~\cite{GFT} where a continuum-geometric phase could emerge through a transition to a condensate phase~\cite{GFTCC1}. 
Investigations of the phase diagram of several GFT models in terms of the functional renormalization-group give indications for such a process in terms of IR fixed points~\cite{TGFTFRG,GFTRG}. 
Such a condensate would correspond to a nonperturbative vacuum described by a large number of bosonic GFT quanta which all settle into a common ground state away from the Fock vacuum. The conjecture that a possible condensate phase can be interpreted as a continuum geometry in GFT models of four-dimensional quantum gravity has spurred the development of the GFT-condensate approach to quantum cosmology~\cite{GFTCC2,GFTCC21,GFTCC22} since under certain assumptions a Friedmann-like dynamics of the emergent space has been found~\cite{GFTCC3,GFTCC4,GFTCC5,GFClowspin}.

Despite these successes, the analysis of the phase structure of GFT models with a proper simplicial gravity interpretation, either in the Euclidean or Lorentzian sector, remains an open problem.
In particular, it would be important to settle the question whether a condensate phase indeed exists in such models.  

In this article we tackle this issue in terms of mean-field techniques. 
Inspired by the phenomenological perspective of Laudau-Ginzburg mean-field theory designed to describe second-order phase transitions, we explore if hints for a phase transition can be found without going through a nonperturbative analysis like the functional renormalization group methodology. 
We apply Landau-Ginzburg theory to the case of the GFT model for three-dimensional Euclidean quantum gravity, augmented by a Laplace-Beltrami operator, hereafter called dynamical Boulatov model~\cite{Boulatov,dynBoulatov}, and then proceed to a conjugation-invariant rank-$1$ GFT model on $\SL(2,\R)$ with local interaction. 
We check the validity of this mean-field approach in the supposed critical region by means of the Ginzburg criterion and find that it provides a trustable description of a phase transition for the noncompact sector of the model on $\SL(2,\R)$. 
In the case of the dynamical Boulatov model on compact group, it proves insufficient to that end.
There, nonperturbative methods remain necessary to analyze if a phase transition can occur or not. 
In this way, our work can be seen as a first step towards the same analysis of the Lorentzian version of the dynamical Boulatov model and, furthermore,  as a precursor to a more involved treatment of these models by means of the functional renormalization group~\cite{Kopietz,FRG}.

The article is organized as follows: In Section~\ref{LGMFT} we recapitulate Landau-Ginzburg mean-field theory and the essence of the Gaussian approximation. This will serve as a template for the analysis for GFT models in the remainder of our work. For the sake of clarity, we will apply this to each individual model to be analyzed.  In Section~\ref{GFT} we introduce some basics of GFT needed for our analysis and we discuss some relevant peculiarities of phase transitions in GFT in Section~\ref{GFTphasetransitions}. Then we apply the mean-field method to GFT models on a compact domain in Section~\ref{LGMFTGFTcompact}. Firstly, we analyze the relevance of gauge symmetry in the case of a rank-$3$ model on $\SU(2)^3$ with quartic local interaction subject to right, left and right, as well as conjugation invariance in Section~\ref{GAlocalGFT}. 
In a next step, we study the effect of nonlocality in Section~\ref{GAnonlocalGFT} in the case of a rank-$1$ toy model on $\SU(2)$ with a convolution-type of interaction and conjugation invariance in Section~\ref{GAnonlocalGFTtoy}, which serves as a warm-up for the case of the dynamical Boulatov model in Section~\ref{GAnonlocalGFTBoulatov}. There, we again discuss the cases where the field is subject to right, left and right, as well as conjugation invariance to demonstrate the independence of the results from the symmetries imposed. In Section~\ref{GAnoncompact} we treat then a rank-$1$ model on the noncompact group $\SL(2,\R)$ with conjugation invariance and quartic local interaction. 
We conclude in Section~\ref{Discussion} with a summary and discussion of our results.
Relevant details of harmonic analysis on the Lie groups $\SU(2)$ and $\SL(2,\R)$ are supplemented in the Appendices.

\section{Landau theory for group field theory}

The aim of this paper is to understand phase transitions in GFT in terms of the Gaussian approximation as pioneered by Landau and Ginzburg~\cite{Ginzburg:1950sr}. 
Thus, we first recapitulate the general scheme,
then remind on the peculiarities of GFT,
and finally discuss the physical meaning of the application of Gaussian approximation to GFT.

\subsection{Landau's theory of phase transitions and the Gaussian approximation}\label{LGMFT}

In the following, we recapitulate the statistical properties of a scalar field $\gf(\vec{x})$ on $\R^\sd$ at mean-field level. We then introduce the Ginzburg criterion which allows us to test the validity of the mean-field description when studying the critical behavior of the system~\cite{Kopietz,Sachs,Kadanoff,Benedetti}.

The generating functional of all correlation functions
\be
Z[J]\equiv e^{W[J]}=\int [\mathcal{D}\gf] e^{-S[\gf]+\int \d ^\sd x~J\gf}
\ee
with external source $J$ defines the statistical field theory of $\gf$. $W[J]$ is the generating functional for the connected correlation functions and
\be
S=\frac{1}{2}\int \d ^\sd x~\gf(\vec{x})\left(-\Delta+m^2\right)\gf(\vec{x})+\frac{\lambda}{4!}\int \d ^\sd x~\gf(\vec{x})^4
\ee
denotes the bare action. 
The connected (two-point) correlation function $\gt$ is given by
\be
\gt(\vec{x}-\vec{x}')=\frac{\delta^2 W[J]}{\delta J(\vec{x})\delta J(\vec{x}')}\bigg \vert_{J=0}
\ee
depending only on the relative coordinate $\vec{x}-\vec{x}'$ due to translation invariance.

One arrives at \textit{Landau's mean-field approximation} when estimating the functional integral $Z[0]$ through the saddle point method for a uniform field configuration $\gf_0$. It minimizes the bare action, that is, solves
the classical equations of motion obtained from $S[\gf_0]$ without source:
\be\label{vacuum}
\gf_0=0~\text{if}~m^2>0~\text{and}~\gf_0=\pm\sqrt{-\frac{m^2}{\lambda/3!}}~\text{if}~m^2<0.
\ee
In the \textit{Gaussian approximation} quadratic fluctuations around the saddle point are retained in $S[\gf]$. Their correlation function $\gt$ is given by the inverse of 
\be
\delta^2_{\gf}S\big\vert_{\gf_0}=-\Delta+m^2~\text{if}~m^2>0~\text{and}~-\Delta-2m^2~\text{if}~m^2<0.
\ee
Equivalently, one can obtain the correlation function from the classical equation of motion with source term in terms of the linearization $\gf(\vec{x})\to \gf_0+\delta\gf(\vec{x})$ and $J(\vec{x})\to J(\vec{x}) +\delta J(\vec{x})$~\cite{Sachs}. This leads to the differential equation
\be\label{SFTeomfluctuations}
\left(-\Delta+m^2\right)\delta\gf(\vec{x})+\frac{\lambda \gf_0^2}{2}\delta\gf(\vec{x}) = \delta J(\vec{x}),
\ee
which we may solve by means of the Green's function method. 
Using the response relation
\be
\delta\gf(\vec{x})=\int \d^\sd x'~\gt(\vec{x}-\vec{x}')\delta J(\vec{x}'),
\ee
the equation of motion Eq.~(\ref{SFTeomfluctuations}) rewrites as
\be
\left(-\Delta+m^2+\frac{\lambda \gf_0^2}{2}\right) \gt(\vec{x})=\delta(\vec{x}),
\ee
which can be solved in Fourier space and leads to an exponentially decaying function in position space.

Given the structure of the effective propagator, the correlation length $\cl$ is defined by
\be\label{clength}
\cl^{-2} = m^2+\frac{\lambda \gf_0^2}{2} \overset{\text{\eqref{vacuum}}}{=}
\left\{ \begin{array}{rccl}
& m^2  & ,  &  m^2>0 \\
& -2m^2  & ,  &  m^2<0 
\end{array} \right .
\ee
setting the scale beyond which the exponential decay in $||\vec{x}-\vec{x}'||$ sets in. At the second-order phase transition $\cl\to\infty$ and the correlation function obeys a power-law behavior.

A test for the validity of the description of the phase transition in terms of the Gaussian approximation is
to quantify the strength of the field fluctuations relative to the mean-field value in terms of the quantity
\be\label{Q}
Q =\frac{\int_{\cl}d^\sd x~\gt(\vec{x}-\vec{x}')}{\int_{\cl}d^\sd x~\gf_0^2}.
\ee
In general, the approximation is self-consistent and deemed trustworthy if $Q\ll 1$ for large $\cl$, that means fluctuations are small along all scales. This condition is the so-called \textit{Ginzburg criterion}~\cite{Kopietz}. In contrast, the approximation breaks down if fluctuations are large, i.e. $Q\gg 1$, necessitating a nonperturbative treatment instead. On flat space $\R^\sd$, the asymptotic behavior for large $\cl$ is
\be
Q \sim \lambda\cl^{4-\sd}
\ee
from which one deduces the breakdown of the Gaussian approximation for the description of the phase transition below the critical dimension $\sd_c=4$.

\subsection{Group field theory}\label{GFT}
A GFT is a quantum field theory with a group configuration space, a gauge symmetry and combinatorially nonlocal interactions~\cite{GFT}.
The central object is a real- or complex-valued scalar field $\gf$ living on $\rk$ copies of a Lie group $G$. Throughout this paper we consider real fields. As an approach to quantum gravity, $G$ should be the gauge group of general relativity and the corresponding symmetry results in an invariance of the GFT action $S[\gf]$ under the (right) diagonal action of $G$ acting on the fields as
\be\label{invariance}
\gf(g_1,...,g_\rk)=\gf(g_1 h,...,g_\rk h),\quad\forall g_i,h\in G,
\ee
which can be imposed via group averaging.

The action is a sum of kinetic and interaction terms, $S[\gf] = \Sk[\gf]+\Sia[\gf]$. 
Throughout this paper we consider the kinetic part to be
\be
\Sk[\gf]=\frac{1}{2}\int [\d g]^\rk \gf(g_1,...,g_\rk) \left(-\Delta+m^2\right)\gf(g_1,...,g_\rk)
\ee
where $\Delta = \sum_{i=1}^\rk \Delta_i$ is the Laplace-Betrami operator on the configuration space $G^{\times\rk}$. 

Interactions in $\Sia[\gf]$ have to be combinatorial nonlocal in the sense that each argument $g_i$ in a term  $\gf(g_1,...,g_i,...,g_\rk)$ is paired via convolution with exactly one $g_j$ in another term $\gf(g_1,...,g_j,...,g_\rk)$ in the product of fields. 
The simplest example is an interaction of order $\rk+1$ where each field is convoluted once with each other. Thus, it has the combinatorics of a complete graph with $\rk+1$ vertices and can be interpreted as dual to an $\rk$-dimensional simplex. 
The path integral 
\be
Z_{GFT}=\int[\mathcal{D}\gf]
e^{-S[\gf]}
\ee
has then a perturbative expansion indexed by Feynman diagrams dual to gluings of $\rk$-simplices and is thus a generating function of certain cellular complexes.\footnote{Adding a coloring~\cite{Gurau:2011dw} or equivalently a tensorial invariance~\cite{Bonzom:2012bg} these complexes are indeed bijective to abstract simplicial pseudomanifolds~\cite{Gurau:2010iu}. Slight extensions of the theory allow also to cover the richness of combinatorial structures present in loop quantum gravity and spin foam models~\cite{ORT14}.}
For this reason, the rank $\rk$ of the GFT is chosen as the dimension of spacetime $\rk=\std$ in order to provide a path integral for quantum gravity. More precisely, it provides a generating function for spin foams~\cite{SF} which can be considered as a covariant formalism for loop quantum gravity~\cite{LQG,GFTLQG}.

Our goal is now to analyze the effect of the various peculiar properties of GFT, 
that is the Lie group $G$, the rank $\rk$, the group symmetry as well as the combinatorial nonlocality, on phase transitions in the Gaussian approximation and check its validity via the Ginzburg criterion closely following the exposition of Section~\ref{LGMFT}. Before we start, we briefly discuss the physical meaning of such a phase transition in GFT.

\subsection{Phase transition in group field theory}
\label{GFTphasetransitions}

The physical meaning of phase transitions in GFT is a research question in its own. 
Technically, it is rather straightforward to apply Landau theory as outlined in Sec.~\ref{LGMFT} to GFT.
However, the original meaning of the scalar field effectively describing degrees of freedom on physical space in condensed-matter physics does not apply here.
This poses a challenge in particular to the concept of correlation length.

In GFT, spacetime itself is generated as the superposition of geometric configurations in correspondence to discrete geometries in terms of the perturbative expansion of the path integral.
As known from matrix models, physically, the most relevant aspect of a phase transition is then that it may describe the critical subspace in coupling space at which an infinite number of such configurations contribute. Approaching the point of phase transition has then the meaning of a limit to continuum spacetime.\footnote{
In tensor models there are examples where such a discrete-to-continuum phase transition can be made precise and related to the spontaneous breaking of unitary symmetry~\cite{Delepouve:2015hc}. To this end, a description of the tensor model in the intermediate-field representation as a multimatrix model is used and perturbations around the nontrivial matrix vacuum are studied.
}
Complementary, if different phases exist on the critical subspace itself, there should also be phase transitions between these (as for example in matrix models~\cite{AlvarezGaume:1992np}, tensor models~\cite{Bonzom:2015gt,Lionni:2017tk} or (causal) dynamical triangulations~\cite{Ambjorn:2012vc}).

In GFT the meaning of \enquote{correlation length} is completely different to the usual notion in condensed-matter physics. There, the correlation length $\cl$ is the scale beyond which correlation functions $\gt(\vec x -\vec{x}')$ on space $\vec x,\vec{x}'\in \R^\sd$ decay exponentially.
Contrary, the GFT configuration space $G^{\times\std}$ is related to parallel transports of the gravitational field through the $\std$ boundaries of a simplicial building block of $\std$-dimensional spacetime. Parallel transports capture the curvature of spacetime geometry.
Thus, a distance on this space describes, roughly speaking, a difference in local curvature. The correlation length describes then the difference of modes with respect to local curvature.

Applying Landau theory to GFT, we consider here phase transitions characterized by arbitrary large correlation length on group space. At this point, fluctuations of arbitrary different group variables, that is parallel transports of the gravitational field, contribute equally to the dynamics.
This is the same physical setting as investigated by functional renormalization group techniques~\cite{TGFTFRG}. 
However, the relation to the discrete-to-continuum limit of GFT or tensor models is not obvious.
From the physical perspective of the discrete geometries, another possibility is that such a phase transition should be characterized by arbitrary large fluctuations in GFT momentum space given by group representations since these are the eigenmodes of length, area or volume operators of such geometries~\cite{Rovelli:1995gq}. While this has been explored in spin foam models~\cite{Steinhaus:2018wp}, the usual GFT propagator does not allow for such a notion of correlation length. 

Even with a correlation length on GFT configuration space there remain some ambiguities. 
Throughout this work we use \eqref{clength} as a definition for $\cl$ as we consider GFT with the standard kinetic term $\Sk[\varphi]$ with Laplacian.
However, for a compact group $G$ with compactness scale $a$, correlations can only decay for geodesic distances between $\cl$ and $a$ such that quantities like Ginzburg's measure for fluctuations \eqref{Q} applied to GFT,
\be\label{relativefluctuations}
Q=\frac{\prod_{i=1}^{\std}\int_{\cl_i} \d g_i ~\gt(g_1,...,g_\std)}{\prod_{i=1}^{\std}\int_{\cl_i}\d g_i \gf_0^2}
\ee
are meaningful only for $\cl$ large but smaller than $a$ (cf.~\cite{Benedetti}).
Furthermore, we integrate all single copies $i=1,...,\std$ of $G$ up to $\cl_i=\cl$ (like on $\R^\sd$ one integrates over a $\sd$-cube with edge length $\cl$~\cite{Kopietz}).
While only a full physical theory of phase transitions in GFT can justify these choices eventually, our Landau analysis already clarifies for the first time various aspects of such transitions through the very necessity to consider the notion of correlation length in GFT.

\section{GFTs on a compact domain in the Gaussian approximation}\label{LGMFTGFTcompact}

In this Section, we firstly discuss a rank-$3$ GFT on $\SU(2)^3$ with an ordinary quartic local interaction with right, left and right as well as conjugation invariance in the Gaussian approximation. Right invariance is the standard symmetry in GFT as explained in Section~\ref{GFT}. If an additional left invariance is imposed, the field domain can be related to the space of homogeneous $2$-geometries, as shown in Ref.~\cite{GFCExample}, and the fixing to conjugation invariance then gives a special case of this scenario. 

Secondly, we explore the effect of nonlocality of interactions in two cases. 
The first is a rank-$1$ toy model on $\SU(2)$ endowed with conjugation invariance with a convolution-type of interaction.\footnote{We want to thank E. Livine for suggesting to us to study this exemplary toy model prior to the more involved case of the dynamical Boulatov model.}
This model shows already the essential features of nonlocality.
In this way it sets the stage for the analysis of the more relevant dynamical Boulatov model in Section~\ref{GAnonlocalGFTBoulatov}.

\subsection{A rank-3 model with a quartic local interaction}\label{GAlocalGFT}

We start off with a local GFT model for a real-valued field $\gf$ living on three copies of the Lie group $G=\textrm{SU}(2)$ subject to different types of invariance as defined below. The model has a quartic local interaction given by
\ba\label{localGFTaction}
S[\gf] &=& \Sk[\gf] +\frac{\lambda}{4!}\int (\d g)^3\gf(g_1,g_2,g_3)^4.
\ea

Minimization of this functional leads to
\begin{align}\label{GAlocaleom}
(-\Delta+m^2)\gf(g_1,g_2,g_3)+\frac{\lambda}{3!}\gf(g_1,g_2,g_3)^3= 0.
\end{align}
In the mean-field approximation, for uniform field configurations it is simply solved by
\be\label{gftvacuum}
\gf_0=0~\text{if}~m^2>0~\text{and}~\gf_0=\pm\sqrt{-\frac{m^2}{\lambda/3!}}~\text{for}~m^2<0.
\ee
In the Gaussian approximation, one considers fluctuations around this background. 
Inserting $\gf\to\gf_0+\delta\gf$ and $J \to J+\delta J$ in Eq.~(\ref{GAlocaleom}) with additional source $J$ and keeping terms to linear order in $\delta\gf$ we find
\be
\left(-\Delta+m^2 +\frac{\lambda\gf_{0}^2}{2!}\right) \delta\gf(g_1,g_2,g_3)= \delta J(g_1,g_2,g_3).
\ee
We solve this equation using the Green's function method.
To this aim, we introduce the response relation for the group field
\begin{multline}\label{responserelation}
\delta\gf(g_1,g_2,g_3)=\\\int (\d h)^3~\gt(g_1 h_1^{-1},g_2 h_2^{-1},g_3 h_3^{-1})\delta J(h_1,h_2,h_3),
\end{multline}
This leads to 
\be\label{GAlocalGreenseqn}
\left(-\Delta+m^2+\frac{\lambda}{2!}\gf_0^2\right)\gt(g_1,g_2,g_3)=\delta(g_1,g_2,g_3),
\ee
which we solve in the spin representation in the next subsections. 
For this, we exploit the fact that the $\delta$-function can be expanded in terms of group characters $\chi^j$ for each representation labeled by half integers $j\in\N/2$, i.e.,
\be
\delta(g_1,g_2,g_3)=\sum_{j_1,j_2,j_3}\prod_{i=1}^3{d_{j_i}}\chi^{j_{i}}(g_i).
\ee
For details regarding the Fourier decomposition on $\SU(2)$, we refer to Appendix~\ref{harmonicanalysissu2}.

\subsubsection{Mono-invariance}\label{GAlocalGFTmono}

At first, we consider GFT with invariance under the right diagonal action of the group, i.e.
\be
\gf(g_1,g_2,g_3)=\gf(g_1 r,g_2 r,g_3r),~\forall g_i, r \in \SU(2)
\ee
which is imposed via group averaging. Hence, the field may be decomposed as
\begin{multline}
\gf(g_1,g_2,g_3)=\\\sum_{m_i,n_i, j_i}\hat{\gf}^{j_1 j_2 j_3}_{m_1 n_1 m_2 n_2 m_3 n_3}\int \d r \prod_{i=1}^3 {d_{j_i}}D_{m_i n_i}^{j_i}(g_i r)\\
=\sum_{m_i,\alpha_i, j_i}\hat{\gf}^{j_1 j_2 j_3}_{m_1 m_2 m_3}
\begin{pmatrix}
j_1 & j_2 & j_3 \\
\alpha_1 & \alpha_2 & \alpha_3 
\end{pmatrix}
\prod_{i=1}^3{d_{j_i}}D^{j_i}_{m_i,\alpha_i}(g_i),
\end{multline}
in terms of $3j$ symbols $\begin{pmatrix}
j_1 & j_2 & j_3 \\
\alpha_1 & \alpha_2 & \alpha_3 
\end{pmatrix}$ and with modes
\be
\hat{\gf}^{j_{1}j_{2}j_{3}}_{m_{1}m_{2}m_{3}}=\sum_{n_1,n_2,n_3}\hat{\gf}^{j_{1}j_{2}j_{3}}_{m_1 n_1 m_2 n_2 m_3 n_3}
\overline{\begin{pmatrix}
j_1 & j_2 & j_3\\
n_1 & n_2 & n_3 
\end{pmatrix}}.
\ee
With this symmetry imposed, for $m^2<0$ the solution to Eq.~(\ref{GAlocalGreenseqn}) reads as
\begin{multline}\label{GAlocalGreensfctmono}
\gt(g_1,g_2,g_3)=\\\sum_{m_i,\alpha_i, j_i}\gth^{j_1 j_2 j_3}_{m_1 m_2 m_3}
\begin{pmatrix}
j_1 & j_2 & j_3 \\
\alpha_1 & \alpha_2 & \alpha_3 
\end{pmatrix}\prod_{i=1}^3{d_{j_i}}D^{j_i}_{m_i,\alpha_i}(g_i)
\end{multline}
where the Fourier coefficients are given by
\be
\gth^{j_{1}j_{2}j_{3}}_{m_{1}m_{2}m_{3}}=\frac{\overline{\begin{pmatrix}
j_1 & j_2 & j_3 \\
m_1 & m_2 & m_3 
\end{pmatrix}}}{\sum_{i=1}^3 j_{i}(j_{i}+1) - 2 m^2}.
\ee

To evaluate the strength of the fluctuations relative to the average field in the supposed region of criticality, we have to compute Eq.~(\ref{relativefluctuations}) for large $\cl$. However, due to compactness of $\SU(2)$ it does not make sense to 
consider $\cl>\pi$ as the compactness scale $a=\pi$ is the maximal possible geodesic distance. 
Thus, we are interested in the \enquote{asymptotic} behavior of $Q$ for large $\cl<a$, that is $\cl$ close to $\pi$.
For this reason it is sufficient to compute the integrals at first simply over the entire $\SU(2)^3$-domain. If we integrate Eq.~(\ref{GAlocalGreensfctmono}) in this way and use the orthogonality relation of the Wigner matrices for each $\SU(2)$-integration, 
we observe that only the zero-mode
$\gth^{000}_{000}=1/2|m^2|$
will contribute to $Q$. 
Since all modes $j>0$ yield continuous oscillations which are zero at $\cl=\pi$, the part of the zero mode is indeed dominant for $\cl$ close to $\pi$.
For the zero mode we can then perform the integration up to $\cl<a$ exactly and find
\be
Q \sim \frac1{-2m^2} \frac1{\gf_0^2} = \frac{\lambda}{3} \cl^4. 
\ee
For large (but smaller than $a$) correlation lengths $\cl^{2}= - m^{-2}/2$ this expression becomes large, indicating the invalidation of the Gaussian approximation in the region of expected phase transition.
In fact, for given $\cl<a$ there are always bare couplings $\lambda\ll a^{-4}$ such that $Q\ll 1$ despite being a power function in $\cl$.
For $a=\pi$ this is the case for $\cl\ll 10^{-2}$, and this value becomes even smaller for larger compactness scales $a$.
Of course, the actual value of the coupling could only be determined by experiment. 
However, the very concept of second-order phase transitions relies on the possibility of correlation lengths $\cl$ to become very large in a physical sense (though described mathematically by asymptotics, physically it is sufficient if they are much larger than the fluctuations around the ground states of the different phases).
Thus, if there is such a phase transition on a compact space, then $\cl$ and as a consequence $Q$ becomes very large indicating the breakdown of the Gaussian approximation.

\subsubsection{Bi-invariance}\label{GAlocalGFTbi}

As a second case, we impose invariance with respect to left and right diagonal action of the group on the group field, i.e.
\be
\gf(g_1,g_2,g_3)=\gf(lg_1 r,lg_2 r,lg_3 r),~\forall g_i, l, r \in \SU(2),
\ee
implemented via group averaging. Hence, the field may be decomposed as
\begin{multline}
\gf(g_1,g_2,g_3)=\\\sum_{m_i,n_i, j_i}\hat{\gf}^{j_1 j_2 j_3}_{m_1 n_1 m_2 n_2 m_3 n_3}\int \d l\int \d r \prod_{i=1}^3 {d_{j_i}}D_{m_i n_i}^{j_i}(lg_i r)=\\
\sum_{\alpha_i,\beta_i, j_i}\hat{\gf}^{j_1 j_2 j_3}\overline{\begin{pmatrix}
j_1 & j_2 & j_3 \\
\alpha_1 & \alpha_2 & \alpha_3 
\end{pmatrix}}\begin{pmatrix}
j_1 & j_2 & j_3 \\
\beta_1 & \beta_2 & \beta_3 
\end{pmatrix} \prod_{i=1}^3 {d_{j_i}}D^{j_i}_{\alpha_i,\beta_i}(g_i) \\
=\sum_{j_1,j_2,j_3}\hat{\gf}^{j_1 j_2 j_3}\int \d h\prod_{i=1}^3{d_{j_i}}\chi^{j_i}(g_i h), 
\end{multline}
where 
\be
\hat{\gf}^{j_{1}j_{2}j_{3}}=\hat{\gf}^{j_{1}j_{2}j_{3}}_{m_1 n_1 m_2 n_2 m_3 n_3}\begin{pmatrix}
j_1 & j_2 & j_3 \\
m_1 & m_2 & m_3 
\end{pmatrix}
\overline{\begin{pmatrix}
j_1 & j_2 & j_3 \\
n_1 & n_2 & n_3 
\end{pmatrix}}.
\ee
With this symmetry imposed, for the sector $m^2<0$ the solution to Eq.~(\ref{GAlocalGreenseqn}) reads as
\begin{multline}\label{GAlocalGreensfctbi}
\gt(g_1,g_2,g_3)=\sum_{j_1,j_2,j_3}\gth^{j_{1} j_{2} j_{3}}\int \d h\prod_{i=1}^3{d_{j_i}}\chi^{j_{i}}(g_ih),
\end{multline}
where the Fourier coefficients are given by 
\be
\gth^{j_{1}j_{2}j_{3}}=\frac{1}{\sum_{i=1}^{3} j_{i}(j_{i}+1) - 2m^2}.
\ee

To evaluate $Q$ in this case, we use the same argument as in the previous subsection. The only difference is that we employ the orthogonality of the characters for each $\SU(2)$-integration to find again that only the zero-mode will contribute when integrating Eq.~(\ref{GAlocalGreensfctbi}) over $\SU(2)^3$. Again we find that the zero-mode thus dominates for $\cl$ large (but smaller than $\pi$) where we obtain
\be
Q \sim \frac{\lambda}{3} \cl^4,
\ee 
indicating the invalidation of the Gaussian approximation in this region.

\subsubsection{Conjugation invariance}\label{GAlocalGFTcon}

Now we consider the case where the field is subject to conjugation invariance, i.e.
\be
\gf(g_1,g_2,g_3)=\gf(kg_1k^{-1},kg_2k^{-1},kg_3k^{-1}),
\ee
which holds for all $g_i$ and $k$ in $\SU(2)$. Hence, $\gf$ is a central function on the domain and can be decomposed in terms of characters, so we write
\be
\gf(g_1,g_2,g_3)=\sum_{j_1,j_2,j_3}\hat{\gf}^{j_{1}j_{2}j_{3}}\prod_{i=1}^3{d_{j_i}}\chi^{j_{i}}(g_i).
\ee
With this symmetry, for $m^2<0$ the solution to Eq.~(\ref{GAlocalGreenseqn}) is
\be
\gt(g_1,g_2,g_3)=\sum_{j_1,j_2,j_3}\gth^{j_{1} j_{2} j_{3}}\prod_{i=1}^3{d_{j_i}}\chi^{j_{i}}(g_i),
\ee
where the Fourier coefficients are again
\be\label{proplocalconj}
\gth^{j_{1} j_{2} j_{3}}=\frac{1}{\sum_{i=3}^3j_i(j_i+1) - 2 m^2}.
\ee

The computation of $Q$ follows along the lines of the previous subsections, leading to
\be
Q  \sim \frac{\lambda}{3} \cl^4.
\ee
Again, this entails $Q\gg1$ in the supposedly critical region. 

We conclude that the Gaussian approximation does not provide a trustable description of a phase transition for the present model subject to the different symmetries.
Furthermore, though we have chosen rank $n=3$ here, it is obvious from the calculations that the result generalizes to arbitrary rank $n$.

The peculiar form of $Q$ is similar to that found for a scalar field with a quartic local interaction on $S^d$ in Ref.~\cite{Benedetti}. There it is furthermore demonstrated through a functional renormalization group analysis that the $\mathbb{Z}_2$-symmetry is always restored in the IR and no phase transition takes place. Such a result might also be found for the models considered here.
However, their full nonperturbative analysis is beyond the scope of this article and will be treated elsewhere.

\subsection{Models with a quartic nonlocal interaction}\label{GAnonlocalGFT}

Now we explore the effect of combinatorial nonlocality on the validity of the Gaussian approximation.
To this end, we consider two models with nonlocal quartic interactions, first a rank-$1$ toy model and second the dynamical Boulatov model. We find similar results for the Ginzburg criterion as for the local model in the preceding Sec.~\ref{GAlocalGFT}.

\subsubsection{Rank-$1$ toy model}\label{GAnonlocalGFTtoy}

It is possible to mimic the nonlocal pairing of field arguments in GFT already for a field  with single argument in terms of a noncommutative convolution product.
Thus, we consider a real-valued field $\gf$ on one copy of $G=\SU(2)$ which is subject to conjugation invariance 
and has dynamics given by the action
\be
S[\gf]= \Sk[\gf] + \frac{\lambda}{4!}\int \d g [\gf\star\gf\star\gf\star\gf](g)
\ee
wherein the convolution product $\star$ is defined via
\be
[\gf\star\gf](g)=\int \d h~\gf(h)\gf(gh^{-1})
\ee
such that the quartic convolution expands into
\begin{multline}
[\gf\star\gf\star\gf\star\gf](g)=\\\int \d h \int\d k \int\d l~\gf(h)\gf(kh^{-1})\gf(lk^{-1})\gf(gl^{-1}).
\end{multline}
Such an interaction captures already the essential aspects of combinatorial nonlocality.

In a first step, we again compute the equation of motion, given by
\ba\label{GAnonlocaltoyeom}
0 &=& (-\Delta+m^2)\gf(g)+\frac{\lambda}{4!}\int \d h\int \d k \int\d l \\
&& \biggl(\gf(kh^{-1})\gf(lk^{-1})\gf(gl^{-1}) +\gf(h^{-1}k)\gf(lk^{-1})\gf(gl^{-1}) \nonumber\\
&& +\gf(h^{-1}k)\gf(k^{-1}l)\gf(gl^{-1}) +\gf(h^{-1}k)\gf(k^{-1}l)\gf(l^{-1}g)\biggr).\nonumber
\ea
For uniform field configurations $\gf_0$ the nonlocality is washed away and the solution is the same as in the local case,
\be
\gf_0=0~\text{if}~m^2>0~\text{and}~\gf_0=\pm\sqrt{-\frac{m^2}{\lambda/3!}}~\text{for}~m^2<0.
\ee
In the Gaussian approximation, however, the nonlocality is retained to a certain degree. 
To show this, we linearize Eq.~(\ref{GAnonlocaltoyeom}) with additional source $J$ via the insertion $\gf\to\gf_0+\delta\gf$ and $J\to J+ \delta J$ while only keeping terms up to linear order in $\delta\gf$. 
For $m^2<0$, this leads to the following integro-differential equation
\be
(-\Delta+m^2)\delta\gf(g)-3 m^2\int \d l~\delta\gf(gl^{-1})
= \delta J(g),
\ee
where the integral term is actually constant due to the properties of the Haar measure. We tackle it using the Green's function method and to this aim introduce the response relation
\be
\delta\gf(g)=\int \d h~ \gt(gh^{-1})\delta J(h).
\ee 
With this we obtain
\be\label{GAnonlocaltoyGreenseqn}
(-\Delta+m^2)\gt(g)-3m^2\int \d l~\gt(gl^{-1})=\delta(g),
\ee
which we solve in Fourier space.
Because of conjugation invariance the correlation function decomposes into
\be\label{Fourierdecomp}
\gt(g)=\sum_j \gth^j{d_j}\chi^{j}(g).
\ee
Using this and the orthogonality relation for the characters (see Appendix~\ref{harmonicanalysissu2}), the integral in Eq.~(\ref{GAnonlocaltoyGreenseqn}) simply contributes a zero-mode $\gth^0$ such that the solution is
\be\label{FCCkonv}
\gth^j=\frac{1}{j(j+1)+m^2-3 m^2\frac{\delta_{j 0}}{d_j}}.
\ee
Hence, in the Gaussian approximation the correlation function obtains a mild modification due to the nonlocality of the interaction.
Comparing to the local case \eqref{proplocalconj}, the zero mode is the same while for modes $j>0$ there is a mass term $m^2$ instead of $-2 m^2 = 2|m^2|$.
Still, the argument for the dominance of the zero mode applies such that we find again the large-$\cl$ behavior 
\be\label{Qnonlocal}
Q \sim \frac{\lambda}{3} \cl^4 .
\ee
Oscillations are stronger by a factor 2 and have opposite sign as compared to the local case, but they remain irrelevant at large $\cl$.

\subsubsection{Dynamical Boulatov model}\label{GAnonlocalGFTBoulatov}

The dynamical Boulatov model~\cite{Boulatov} is a GFT with real-valued field $\gf$ on three copies of $\SU(2)$ with a simplicial quartic interaction
\be\label{Boulatovaction}
S\left[\gf\right] = \Sk[\gf]+\frac{\lambda}{4!}\int(\d g)^{6}~\gf_{123}\gf_{145}\gf_{256}\gf_{364}
\ee
where we abbreviate $\gf_{ijk}\equiv\gf(g_i,g_j,g_k)$ from now on.
The action is endowed with an invariance with respect to the right diagonal action of the group $\SU(2)$, usually imposed via group averaging.\footnote{In the original definition of the Boulatov model there is furthermore an invariance with respect to cyclic permutations of field arguments~\cite{Boulatov}. This symmetry will not play a role in the following considerations.} 
The action is constructed such that the perturbative expansion of the generating
functional around the Fock vacuum is equivalent to the Ponzano-Regge spin foam model~\citep{PonzanoRegge,SF} which provides a discrete version of the path integral for three-dimensional Euclidean quantum gravity.

As in the other cases, we first compute the equation of motion, 
\be\label{GABoulatoveom}
(-\Delta+m^2)\gf_{123}+\frac{\lambda}{3!}\int \d g_4 \d g_5 \d g_6 \gf_{146}\gf_{526}\gf_{543}=0.
\ee
The projection onto uniform field configurations $\gf_0$ is not sensitive to the combinatorial nonlocality. Thus, it is solved again by
\be
\gf_0=0~\text{if}~m^2>0~\text{and}~\gf_0=\pm\sqrt{-\frac{m^2}{\lambda/3!}}~\text{for}~m^2<0.
\ee
Turning to the Gaussian approximation, the effect of the nonlocality appears for small deviations around this constant background. To see this, we linearize Eq.~(\ref{GABoulatoveom}) with additional source $J$ via inserting $\gf\to\gf_0+\delta\gf$ and $J\to J+\delta J$ yielding
\begin{multline}
(-\Delta+m^2)\delta\gf_{123}\\ +\frac{\lambda}{3!} \gf_0^2\int \d g_4 \d g_5 \d g_6 (\delta\gf_{146} + \delta\gf_{526} + \delta\gf_{543}) = \delta J_{123}.
\end{multline}
We solve this integro-differential equation using again the response relation \eqref{responserelation} such that
\begin{multline}\label{GABoulatovGreenseqn}
\biggl[-\Delta+m^2+ \frac{\lambda}{3!} \gf_0^2\left(\int \d g_2 \d g_3 + \int \d g_1 \d g_3 + \int \d g_1 \d g_2 \right) \biggr] \\ 
\gt(g_1 h^{-1}_1,g_2 h^{-1}_2,g_3 h^{-1}_3) = \prod_{i=1}^3\delta(g_i h_i^{-1}).
\end{multline}
In the following, we solve this equation in Fourier space for three types of invariance. To deal with the integral kernel, we use the orthogonality relation of the Wigner matrices and characters, see Appendix~\ref{harmonicanalysissu2}, just in the same way as in Section~\ref{GAnonlocalGFTtoy}.

\paragraph*{\small\textit{~~~~~~~~~~~~~~~~~~~~~~~2.1~~~Mono-invariance}}\label{GAnonlocalGFTBoulatovmono}

For field configurations simply endowed with the invariance with respect to the right diagonal action, the correlation function is
\begin{multline}
\gt(g_1,g_2,g_3)=\\\sum_{m_i,\alpha_i, j_i}\gth^{j_1 j_2 j_3}_{m_1 m_2 m_3}
\begin{pmatrix}
j_1 & j_2 & j_3 \\
\alpha_1 & \alpha_2 & \alpha_3 
\end{pmatrix}\prod_{i=1}^3{d_{j_i}}D^{j_i}_{m_i,\alpha_i}(g_i).
\end{multline}
The Fourier coefficients are given by
\be
\gth^{j_{1}j_{2}j_{3}}_{m_{1}m_{2}m_{3}}=\frac{\overline{\begin{pmatrix}
j_1 & j_2 & j_3 \\
m_1 & m_2 & m_3 
\end{pmatrix}}}{\sum_{i=1}^3 j_{i}(j_{i}+1)+m^2-3m^2 A},
\ee
with 
\ba
A &=& \sum_{i<k}
\frac{\delta_{j_i 0}}{d_{j_i}}\delta_{m_{i}0}\delta_{\alpha_{i}0}
\frac{\delta_{j_k 0}}{d_{j_k}}\delta_{m_{k}0}\delta_{\alpha_{k}0} \\
& = &\delta_{j_20}\delta_{j_30}+\delta_{j_10}\delta_{j_20}+\delta_{j_10}\delta_{j_30}.
\ea
This term leads to a mild modification of the correlation function in the Gaussian approximation similar to the toy model above which is due to the nonlocality of the interaction.
Hence, we find also the same result for the relative fluctuations $Q$ up to the numerical factors given by $3A$ which slightly modify the amplitude of $j_i>0$ mode oscillations but do not influence the large-$\cl$ behavior $Q\sim\lambda \cl^4/12$ due to the zero mode.

\paragraph*{\small\textit{~~~~~~~~~~~~~~~~~~~~~~~~~2.2~~~Bi-invariance}}\label{GAnonlocalGFTBoulatovbi}

For field configurations endowed with the invariance with respect to the left and right diagonal action, the correlator is given by
\be
\gt(g_1,g_2,g_3)=\sum_{j_1,j_2,j_3}\gth^{j_{1} j_{2} j_{3}}\int \d h\prod_{i=1}^3{d_{j_i}}\chi^{j_{i}}(g_ih).
\ee
Its Fourier coefficients are 
\be
\gth^{j_{1}j_{2}j_{3}}=\frac{1}{\sum_{i=1}^{3} j_{i}(j_{i}+1)+m^2-3m^2B},
\ee
with
\be
B = 3\frac{\delta_{j_1 0}}{d_{j_1}}\frac{\delta_{j_2 0}}{d_{j_2}}\frac{\delta_{j_3 0}}{d_{j_3}}.
\ee
Due to the nonlocality of the interaction, the last term again gives a mild modification of the correlation function in the Gaussian approximation. Its particular form varies from that of the previous case due to the different symmetry imposed onto the field.
Still the qualitative behavior is the same and $Q$ has the same asymptotics as the previous case.

\paragraph*{\small\textit{~~~~~~~~~~~~~~~~~~2.3~~~Conjugation invariance}}\label{GAnonlocalGFTBoulatovcon}

For field configurations subject to conjugation invariance, the solution to Eq.~(\ref{GABoulatovGreenseqn}) expands as 
\be
\gt(g_1,g_2,g_3)=\sum_{j_1,j_2,j_3}\gth^{j_{1} j_{2} j_{3}}\prod_{i=1}^3{d_{j_i}}\chi^{j_{i}}(g_i),
\ee
with Fourier coefficients given by
\be
\gth^{j_{1} j_{2} j_{3}}=\frac{1}{\sum_{i=3}^3j_i(j_i+1)+m^2-3m^2C},
\ee
and
\be
C = \left(\frac{\delta_{j_2 0}}{d_{j_2}}\frac{\delta_{j_3 0}}{d_{j_3}}+\frac{\delta_{j_1 0}}{d_{j_1}}\frac{\delta_{j_2 0}}{d_{j_2}}+\frac{\delta_{j_1 0}}{d_{j_1}}\frac{\delta_{j_3 0}}{d_{j_3}}\right).
\ee
Again, we find a mild modification of the correlation function in the Gaussian approximation due to the nonlocality of the interaction. 
Qualitatively, it yields the same result as in the other cases.

We may conclude that, following Landau's strategy, the nonlocal interactions treated here have no relevant effect on the singular behavior of $Q$. Hence, the Gaussian approximation cannot be trusted to give a valid description of a phase transition for these models. Nonperturbative methods have to be applied to settle the question if a phase transition can take place for these. 

This result generalizes not only to simplicial interactions of different rank but also to other types of nonlocality such as tensor-invariant interactions (as studied for example in~\cite{TGFTFRG}). The reason is that they all lead to integro-differential equations of the type \eqref{GABoulatovGreenseqn} differing only in the specific structure of integrations. For tensor-invariant interactions one has for example terms with different number of integrations. But the result is only a different specific form of $\delta_{j0}$ terms in the modification of the representation-space propagator (like the terms $A$, $B$ and $C$ above).
These are responsible only for the slight modification of higher-mode oscillations but do not alter the dominant zero-mode contribution.

\

Concluding this section on GFT on a compact group, it is important to emphasize once more that it is solely the zero mode of fields on compact manifolds which causes the breakdown of the Gaussian approximation as a description for the theory at phase transition.
From the perspective of loop quantum gravity one might alternatively be interested in a modified GFT excluding these zero modes.
This is because edges with variable $j=0$ are equivalent to no edge at all in the construction of the kinematical Hilbert space in terms of cylindrical functions on embedded graphs.\footnote{We refer to Ref.~\cite{GFTLQG} where subtle differences between the kinematical Hilbert spaces of LQG and GFT are discussed in detail.}
For such a modified GFT, the result of Landau-Ginzburg theory is possibly completely the opposite, that is, the Gaussian approximation could be valid.

\section{GFT on a noncompact domain in the Gaussian approximation}\label{GAnoncompact}

To overcome the issue of large Gaussian fluctuations in GFT on compact configuration space the natural consequence is to consider GFT with noncompact groups.
From a quantum-gravity perspective they are also more interesting since they provide models with Lorentzian signature.
However, the application of Landau theory to 
the Lorentzian dynamical Boulatov model is not straightforward. 
A geometric GFT model for Lorentzian spacetimes in $3d$ has to be based on three copies of the Lie group $\SL(2,\R)$. 
Due to noncompactness already the bare GFT action in Lorentzian signature is well defined only upon regularization. 
This is because the imposition of the right invariance yields spurious integrations over at least one copy of $\SL(2,\R)$ leading to group volume divergences~\cite{GFTCC21}. 
Next to the increased degree of difficulty due to the intricacies of the representation theory of $\SL(2,\R)$, not to mention the handling of the tensor product decomposition for a rank-$3$ model, the volume divergences are the main reason why we devote our attention to a simplified scenario here. 

In the following, we discuss a rank-$1$ toy model on $\SL(2,\R)$ with a local quartic interaction in the Gaussian approximation. 
We also restrict our analysis to conjugation-invariant fields, which simplifies the harmonic analysis. 
This model has no obvious geometric interpretation but, due to the locality of the interaction, it is free of the aforementioned divergences. 
In this way, the following work serves as a first step towards the analysis of geometric models, focusing on the influence of the noncompact domain onto the critical behavior.

\subsection{A rank-1 toy model on $\SL(2,\R)$ with a quartic local interaction}\label{GAnoncompactlocal}

In the following, we consider a GFT model for a real-valued field $\gf$ living on one copy of $G=\SL(2,\R)$ which is subject to conjugation invariance. Its dynamics are defined by
\be
S[\gf]= \Sk[\gf]
+\frac{\lambda}{4!}\int \d g\, \gf(g)^4,
\ee
wherein $\d g$ denotes the Haar measure on $\text{SL}(2,\mathbb{R})$. In a first step, we again compute the equation of motion, given by
\be\label{GAnoncompactlocaltoyeom}
\left(-\Delta+m^2\right)\gf(g)+\frac{\lambda}{3!}\gf(g)^3=0.
\ee
In the mean-field approximation, for uniform field configurations it is solved by
\be
\gf_0=0~\text{if}~m^2>0~\text{and}~\gf_0=\pm\sqrt{-\frac{m^2}{\lambda/3!}}~\text{for}~m^2<0.
\ee
We arrive at the Gaussian approximation by linearizing Eq.~(\ref{GAnoncompactlocaltoyeom}) with additional source $J$ via the insertion $\gf\to\gf_0+\delta\gf$ and $J\to J+\delta J$ while only keeping terms up to linear order in $\delta\gf$. This leads to the differential equation 
\be
\left(-\Delta+m^2\right)\delta\gf(g)+\frac{\lambda}{2!}\gf_0^2\delta\gf(g)=\delta J(g),
\ee
which we solve once again via the Green's function method leading to
\be\label{GAnoncompactlocaltoyGreenseqn}
\left(-\Delta+m^2+\frac{\lambda}{2!}\gf_0^2\right)\gt(g)=\delta(g)
\ee
which we solve in representation space.

To solve this differential equation, we explain briefly the relevant features of $\SL(2,\R)$ as well as harmonic analysis thereon and refer to Appendix~\ref{SLRandreptheory} for further details. 
The group $\SL(2,\R)$ has two Cartan subgroups, a compact one corresponding to rotations
\be
    H_0=\left\{u_{\theta}=
    \begin{pmatrix}
\cos\theta & \sin\theta \\
-\sin\theta & \cos\theta 
\end{pmatrix},~ 0\leq \theta \leq 2\pi\right\}.
\ee
and a noncompact one corresponding to boosts,
\be
    H_{1}=\left\{\pm a_{t}=
    \begin{pmatrix}
\pm e^t & 0 \\
0 & \pm e^{-t} 
\end{pmatrix},~ t\in \R\right\}.
\ee
A regular group element can be conjugated to either one or the other. Using group averaging arguments~\cite{FreidelLivine,Marolf}, it follows that a conjugation-invariant field defined on $\SL(2,\R)$ is either supported on the classes of group elements conjugated to $H_0$ or to $H_1$.\footnote{These two sectors cannot be mapped into one another which can be interpreted as a superselection rule~\cite{Marolf}.} 
We call these conjugation classes $G_0$ and $G_{\pm}$.
Upon averaging over them, the field  depends only on an angle $\theta\in[0,2\pi]$ (parametrizing the compact domain) or $t\in\R$ (parametrizing the noncompact domain), as explained in detail in Appendix~\ref{SLRstuff}. 
Then we have to analyze the Gaussian approximation for such averaged objects separately.

\subsubsection{Gaussian approximation for fields averaged over compact subgroup}

To solve Eq.~(\ref{GAnoncompactlocaltoyGreenseqn}) for fields averaged over $G_0$, we use their decomposition and that of the $\delta$-distribution as explained in Appendix~\ref{SL2RPlancherel}. 
The Green's function decomposes for $m^2<0$ as
\be\label{FourierdecompSL}
\gt(\theta)= \sum_{n=1}^\infty
\frac{n}{4\pi}\left(\gth^+(n) \chi_n^{+}(\theta)+ \gth^-(n) \chi_n^{-}(\theta)\right),
\ee
where the Fourier coefficients for $n=1,2,...$ are
\be\label{propcompact}
\gth(n)\equiv\gth^\pm(n)=\frac{1}{\frac{1-n^2}{4}- 2 m^2}.
\ee

Due to the restriction to the compact direction, the evaluation of the strength of fluctuations in terms of $Q$, given in this case by 
\be
Q=\frac{\int_{\cl}\d \theta\sin^2\theta \,\gt(\theta)}{\int_{\cl}\d \theta\sin^2\theta \,\gf_0^2},
\ee
closely follows our observations for the local GFT models on a compact domain constructed from $\SU(2)$ in Sec.~\ref{GAlocalGFT}. 
Again, due to the compactness of the domain it does not make sense to evaluate the integrals therein for $\cl\to\infty$ but instead only up to $\cl < \pi$. 
We find the same behavior as before for $\SU(2)$, namely
\be
Q \sim \frac{1}{- 2 m^2}\frac{1}{\gf_0^2} = \frac{\lambda}{3}\cl^4,
\ee 
leading to the invalidation of the Gaussian approximation for large $\cl$.
The dominant behavior stems here from the modes for $n=1$ which play the role of the zero-mode contribution discussed in Section~\ref{GAlocalGFT}. 
Indeed the two modes labeled by $n=1$ are zero modes in the sense that they have zero eigenvalue with respect to the Laplacian.
The contributions for all the other modes can be neglected. 
We may also note that due to the structure and resemblance of the characters $\chi_n^{\pm}(\theta)$ to those of $\SU(2)$, the analysis of a model on the former with a nonlocal convolution-type of interaction will reproduce the same result for $Q$ as for the latter, Eq.~(\ref{Qnonlocal}).

\subsubsection{Gaussian approximation for fields averaged over noncompact subgroups}

For fields which are averaged over $G_{\pm}$, we use the decomposition and that of the $\delta$-distribution expatiated on in Appendix~\ref{SL2RPlancherel} to solve Eq.~(\ref{GAnoncompactlocaltoyGreenseqn}). 
The Green's function decomposes in the sector $m^2<0$ as
\ba
\gt(t) &=&
\int_0^\infty\frac{\d s}{4\pi} \frac{s}{2}\gth(s)\left(\tanh\frac{\pi s}{2}\chi_s^+(t)+\coth\frac{\pi s}{2}\chi_s^{-}(t)\right)\nonumber\\
&&+\sum_{n=1}^\infty
\frac{n}{4\pi}\gth(n)\left(\chi_n^{+}(t)+\chi_n^{-}(t)\right)
\ea
where the Fourier coefficients $\gth(n)$ are as in \eqref{propcompact} and the coefficients of the continuous series are
\be\label{propnoncompact}
\gth(s)\equiv\gth^\pm(s)=\frac{1}{\frac{1+s^2}{4} - 2 m^2}.
\ee
It is possible to obtain exact expressions for the different contributions to the Green's function and thus quantify the behavior of field fluctuations via $Q$. We discuss this in the following and refer for details to Ref.~\cite{APthesis}.

To compute the part of $\gt(t)$ stemming from the continuous series, we use the expression for the $\delta$-distribution in Appendix~\ref{SL2RPlancherel} and compute the contributions of the positive and negative branches, separately. 
To this end, we use the series expansions of $\tanh$ and $\coth$
\ba
\frac{\pi s}{2}\tanh\frac{\pi s}{2} &=&\sum_{n\in 2\mathbb{Z}+1}\frac{s^2}{s^2+n^2} \, ,\\
\frac{\pi s}{2}\coth\frac{\pi s}{2} &=&\sum_{n\in 2\mathbb{Z}}\frac{s^2}{s^2+n^2}
\ea
and apply the residue theorem to compute the integrals. This yields
\begin{widetext}
\ba
C_{\text{cont}}^{+}(t) &=&\frac{1}{4\pi}\int_0^\infty\d s\frac{s}{2}\gth(s)\tanh\frac{\pi s}{2}\chi_{s}^{+}(t)\\ 
&=&\sqrt{\frac{\pi}{2}}\frac{1}{|\sinh t|} \Biggl(-\frac{\pi}{2} e^{-|t|\sqrt{1-8m^2}}\tan \left(\frac{\pi}{2}\sqrt{1-8m^2}\right)
-\sum_{n\in 2\mathbb{Z}+1}\frac{|n|e^{-|n||t|}}{1-n^2-8m^2}\Biggr)
\ea
and
\ba
C_{\text{cont}}^{-}(t) &=& \frac{1}{4\pi}\int_0^\infty\d s\frac{s}{2}\gth(s)\coth\frac{\pi s}{2}\chi_{s}^{-}(t) \\
&=& \sqrt{\frac{\pi}{2}}\frac{\text{sgn}(\lambda_{\pm a_t})}{|\sinh t|}\Biggl(\frac{\pi}{2} e^{-|t|\sqrt{1-8m^2}}\cot \left(\frac{\pi}{2}\sqrt{1-8m^2}\right)
- \sum_{n\in 2\mathbb{Z}\backslash\{0\}}\frac{|n|e^{-|n||t|}}{1-n^2-8m^2}\Biggr).
\ea
\end{widetext}

To compute the part of $\gt(t)$ originating from the discrete series on the noncompact direction, we can proceed as in the previous subsection and write
\be
C_{\text{disc}}^{+}(t)=\frac{1}{2|\sinh t|} \sum_{n=1}^\infty
\frac{e^{-n|t|}}{1-n^2 - 8m^2}
\ee
and
\be
C_{\text{disc}}^{-}(t)=\frac{\text{sgn}(\pm a_t)}{2|\sinh t|} \sum_{n=1}^\infty
\frac{e^{-n|t|}}{1-n^2 - 8m^2} \,.
\ee
The sums over $n$ appearing in each of these expressions converge to sums of hypergeometric functions ${}_{2}F_1$ whose details are not relevant here.

We are now ready to quantify the strength of fluctuations by evaluating
\be
    Q=\frac{\int_{\cl}dt~\sinh^2 t \gt(t)}{\int_{\cl}dt~\sinh^2 t\gf_0^2}.
\ee
We proceed step by step and compute this expression for the individual contributions to the Green's function. To this aim, it is sufficient to estimate $Q$ by looking at the asymptotic behavior of the integrand in the numerator for $t\to\infty$ and integrating it for $\cl\to\infty$ since modes beyond $\cl$ are anyway exponentially suppressed~\cite{Kopietz, Benedetti}.
In the denominator we have to integrate up to finite $\cl$. In this way we obtain for large $\cl$
\be
    Q\sim\lambda \cl^4 e^{-2\cl}.
\ee 
Hence, the Gaussian approximation is valid at large $\cl$ where $Q\ll1$.
Thus, it provides a trustworthy description of a phase transition at which $\cl\to\infty$. 

This result is in agreement with the one obtained for a scalar field with quartic local interaction on the $3d$ hyperboloid $\mathbb{H}^3$~\cite{Benedetti}. We may understand the similarity of the results from the fact that $\SL(2,\R)\cong\mathrm{AdS}^3$ which in turn is diffeomorphic to $\mathbb{H}^{1,2}$.

The form of $Q$ is in stark contrast to the results of the previous sections for fields living on compact domains constructed from $\SU(2)$ and suggests that for a phase transition to occur in the GFT context (and to be visible already at the mean-field level), the noncompactness of the domain is a decisive prerequisite.

\section{Discussion and conclusion}\label{Discussion}
 
The purpose of this article was to investigate the critical behavior of various GFT models with and without geometric interpretation in the Gaussian approximation. This encompassed the analysis of the validity of the mean-field techniques employed to this end. In the following, we list the different models and the respective results. 

$(1)$ With the example of a rank-$3$ model on $\SU(2)^3$ with a quartic local interaction subject to right, left and right as well as conjugation invariance we showed that the mean-field techniques break down at large correlation length $\cl$, irrespective of the symmetries imposed onto the field.

$(2)$ The case of a rank-$1$ model on $\SU(2)$ with a quartic nonlocal interaction of convolution-type subject to conjugation invariance showed that the mean-field techniques seize to be valid at large $\cl$.
Nonlocality does effect higher field modes but without changing the order of magnitude. 
The dominant zero mode responsible for the breakdown of the Gaussian approximation is not altered by nonlocality.

$(3)$ For the dynamical Boulatov model we found the same result as for the nonlocal toy model, only the exact prefactors of higher modes depend on the specific type of nonlocality. 
In this case, we checked also right, left and right as well as conjugation invariance to demonstrate the independence of the result from the symmetries imposed. 
We attribute the failure of the mean-field techniques to the compactness of the field domain used and expect the result to generalize to other nonlocal interactions such as simplicial interactions for different rank or tensor-invariant interactions.

Finally, in $(4)$ we analyzed the critical behavior in the case of a rank-$1$ GFT model on $\SL(2,\R)$ with a quartic local interaction subject to conjugation invariance. To our best knowledge, in spite of its toy model nature, this is the first time a GFT model with Lorentzian signature has been studied in some detail in the literature. We employed group averaging arguments to separately analyze the validity of the mean-field approach for fields averaged over the conjugation classes of the two Cartan subgroups. For the compact direction, we obtained results analogous to the ones found in case $(1)$, whereas for the noncompact direction mean-field techniques continue to be valid in the critical region and can serve as a trustable description of a phase transition. This is ultimately rooted in the noncompactness of the field domain.

In the following, we want to comment on the limitations and possible extensions of our discussion.

Given the breakdown of the mean-field techniques towards the supposedly critical region for the cases $(1)$-$(3)$, the impact of higher order fluctuations should be investigated by means of nonperturbative techniques as for example the functional renormalization group. 
With these it should be possible to decide whether or not a phase transition can occur. 

In this sense, our work can also be seen as a motivation to extend the successful functional renormalization group methodology developed for tensorial GFT~\cite{TGFTFRG} to the realm of simplicial GFT. Notice that even if no evidence for a phase transition towards a condensate phase would be found for such models, this would not mean that GFT-condensate phases (lying at the heart of the condensate cosmology approach) cannot exist for these but simply that they cannot be realized through a phase transition then. It would still perfectly make sense to model nonperturbative vacua of such models by mean-field methods and explore to which mean geometry they would correspond.

Before nonperturbative methods are applied, it could be instructive to go beyond the particular realization of Landau mean-field theory with the Gaussian approximation by relaxing one of the main assumptions of this approach, namely the projection onto uniform field configurations. A starting point of such a study could be the nontrivial (not uniform) global minima of the dynamical Boulatov model for right and left invariant as well as equilateral field configurations found in Ref.~\cite{Boulatovnpertvac}.

Finally, in view of the last part of our work, it would be important to extend the analysis for the locally interacting rank-$1$ toy model on $\SL(2,\R)$ to the rank-$3$ case where only right invariance is imposed. For this, Ref.~\cite{ConradyHnybida} could be useful which collects a variety of facts on the representation theory of $\SU(1,1)$ (which is diffeomorphic to $\SL(2,\R)$). In a second step, a regularization scheme should be introduced to tackle the volume divergences for models with a nonlocal interaction possibly of simplicial type. As an intermediate pedagogical step, a rank-$1$ toy model with a convolution-type of interaction and conjugation invariance should be studied to this end. At the level of the correlator, this would already indicate possible modifications which could be expected for case of the full-blown rank-$3$ model with simplicial interaction and Lorentzian signature, similar as for the case of the dynamical Boulatov model. 
The goal of such considerations would of course be to understand if phase transitions and different phases can actually exist for such a model
and whether these are related to $(2+1)$-dimensional Lorentzian continuum geometries at all. From a larger perspective, this would also allow us to establish contact and compare with the existing literature on $(2+1)$-dimensional Lorentzian loop quantum gravity and spin foam models~\cite{FreidelLivineRovelli,Davids}.

\subsection*{Acknowledgements}
The authors thank S. Carrozza, J. Ben Geloun, E. Livine, and D. Oriti for useful remarks. 

J.T. was supported by the German Academic Exchange Service (DAAD) with funds from the German Federal Ministry of Education and Research (BMBF) and the People Programme (Marie Curie Actions) of the European Union's Seventh Framework Programme (FP7/2007-2013) under REA grant agreement n$^\circ$ 605728 (P.R.I.M.E. - Postdoctoral Researchers International Mobility Experience).

\appendix
\section*{Appendix}

\section{Harmonic analysis on Lie Groups}\label{harmonicanalysisgeneral}

Fourier transformations on flat space can be generalized to semi-simple compact Lie groups and to some extent also to noncompact ones.
One can use irreducible unitary representations $\pi$ to define a transform of a $L^2$-function on the Lie group $G$ to a function $\hat{f}$ on representation space,
\be
\hat{f}(\pi)=\int_G \d g f(g)\pi_{g^{-1}}
\ee
in terms of the Haar measure $\d g$.
If available, the Plancherel inversion formula describes the decomposition of $f$ into such modes,
\be\label{fourierinversion}
f(g)=\int_{\hat{G}}\d \mu(\pi^{\lambda})\tr\left(\hat{f}(\pi^{\lambda})\pi^{\lambda}_g\right)
\ee
where $\hat{G}$ is the unitary dual of $G, i.e.$, $\hat{G}$ is the set of all equivalence classes of irreducible unitary representations of $G$. One can choose a representation $\pi^{\lambda}$ for each class $\lambda$ in $\hat{G}$. The Plancherel measure is denoted by $\d \mu(\pi^{\lambda})$, see Refs.~\cite{Knapp,Gurarie,SL2Rsonst1} for details.

Accordingly, the Plancherel theorem for $L^2$-functions on $G$ is
\be
\int_G\d g|f(g)|^2=\int_{\hat{G}}\d \mu(\pi^{\lambda})||\hat{f}(\pi^{\lambda})||_{HS}^2
\ee
with  $||\hat{f}(\pi^{\lambda})||_{HS}^2=\tr\left(\hat{f}(\pi^{\lambda})\hat{f}(\pi^{\lambda})^{*}\right)$ the Hilbert-Schmidt norm. The direct-integral decomposition of the regular representation $R_g$ for $g\in G$ into the sum of primary components is
\be
R_g\simeq~\mathclap{\displaystyle\int_{\hat{G}}}\mathclap{\textstyle\oplus}~~~\d \mu(\pi^{\lambda})\pi_g^{\lambda}\otimes \id_{d(\pi^{\lambda})},
\ee
where $\id_d$ is the identity on the vector space of dimension given by multiplicity $d=d(\pi^{\lambda})$ which may be finite or infinite~\cite{Gurarie}.

\section{Elements of Fourier analysis on $\SU(2)$, properties of Wigner matrices and characters}\label{harmonicanalysissu2}

On a semi-simple compact Lie group unitary irreducible representations act on finite vector spaces and representations have matrix coefficients~\cite{AnalysisonLieGroups,Biedenharn, Knapp}.
In particular, on $G=\SU(2)$ unitary irreducible representations can be labeled by half integers $j\in\left\{ 0,\frac{1}{2},1,\ldots\right\} $ and the representation spaces have dimension $d_{j}=2j+1$. 
The matrix coefficients are given by the Wigner matrices $D^{j}(g)$. Thus, the Plancherel inversion formula for an $L^2$-function $f$ on $\SU(2)$ takes the form
\ba
f(g)&=&\sum_j \mu(\pi^j) \tr\left(\hat{f}(\pi^j)\pi^j_g\right)\\
&=&\sum_j{d_j}\sum_{m,n=-j}^{j} f^j_{mn}D^j_{mn}(g)
\ea
where $f^j_{mn}$ are the coefficents of the transforms $\hat{f}(\pi^j)$. As a special example, the $\delta$-distribution with transforms $\hat{\delta}(\pi^j)=\id_{d_j}$ for all $j$ is given by
\be
\delta(g)=\sum_j{d_j}\chi^j(g)
\ee
in terms of characters $\chi^{j}(g)\equiv\tr D^{j}(g)$. Tensor product representations are easily obtained from this~\cite{Biedenharn}.

Some relevant properties of Wigner matrices and characters are the following:
\begin{enumerate}
\item Wigner-matrix coefficients form an orthogonal basis in $L^{2}(\SU(2))$ with
\be
\int\d g\,D_{m_{1}n_{1}}^{j_{1}}\left(g\right)\bar{D}_{m_{2}n_{2}}^{j_{2}}\left(g\right)=\frac{1}{d_{j_{1}}}\delta^{j_{1}j_{2}}\delta_{m_{1}m_{2}}\delta_{n_{1}n_{2}}.
\ee
\item These coefficients form a basis of eigenfunctions for
the Laplace-Beltrami operator $-\Delta$, i.e.
\be
-\Delta D_{mn}^{j}\left(g\right)=j\left(j+1\right)D_{mn}^{j}\left(g\right).
\ee
\item The characters are smooth real-valued functions satisfying $\chi^{j}\left(g\right)=\chi^{j}\left(g^{-1}\right)$.
\item For $g_{1},g_{2}\in\SU\left(2\right)$ one has the convolution
relation
\be
\int dh~\chi^{j}(hg_{1})\chi^{l}(g_{2}h)=\frac{\delta_{jl}}{d_{j}}\chi^{j}(g_{2}g_{1}^{-1})
\ee
from which the orthogonality relation $\int dh\chi^{j}(h)\chi^{l}(h)=\delta_{jl}$ is retrieved.
\end{enumerate}

\section{Representations and harmonic analysis on $\SL(2,\R)$}\label{SLRandreptheory}
\subsection{Group structure of $\SL(2,\R)$}\label{SLRstuff}
The noncompact, simple and multiply connected Lie group $G=\SL(2,\R)$ is the group of $2\times 2$ real matrices of determinant $1$, i.e.,
\be
\SL(2,\R)=\left\{
\begin{pmatrix}
a & b \\
c & d\end{pmatrix},~a,b,c,d\in\R, ~ad-bc=1\right\}. 
\ee
Its largest normal subgroup is its centre
$Z=\left\{\pm \mathds{1}\right\}$. The group acts by linear transformation on $\R^2$ while preserving oriented area. The eigenvalues of a matrix $g\in\SL(2,\R)$ are
\be\label{eq:lambda_g}
    \lambda_{g}^\pm =\frac{\tr{(g)}\pm \sqrt{(\tr{(g)})^2-4}}{2}.
\ee
such that elements in $SL(2,\R)$ are classified
according to the following scheme:
\begin{itemize}
    \item If $|\tr{(g)}|<2$, $g$ is called elliptic,
    \item if $|\tr{(g)}|=2$, $g$ is called parabolic,
    \item if $|\tr{(g)}|>2$, $g$ is called hyperbolic.
\end{itemize}

There are different ways to decompose $G$ in terms of three special subgroups:
\begin{itemize}
\item the maximal compact subgroup (isomorphic to $\mathrm{SO}(2)$)
\be\label{eq:H0}
H_0=\left\{u=u_{\theta}=
    \begin{pmatrix}
\cos(\theta) & \sin(\theta) \\
-\sin(\theta) & \cos(\theta) 
\end{pmatrix},~ 0\leq \theta \leq 2\pi\right\},
\ee
\item the upper/lower unipotent subgroups
\be
N=N_{\pm}=\left\{n=
\begin{pmatrix}
1 & \nu \\
0 & 1
\end{pmatrix}~\text{and}~
\begin{pmatrix}
1 & 0 \\
\nu & 1
\end{pmatrix},~\nu\in\R \right\}
\ee
\item and the diagonal group 
\be\label{eq:H1}
    H_1=\left\{a=\pm a_{t}=
    \begin{pmatrix}
\pm e^t & 0 \\
0 & \pm e^{-t} 
\end{pmatrix},~ t\in \R\right\}
\ee
with its positive part, the semigroup
\be
H_{1,+}=\{a_t:t>0\}.
\ee
\end{itemize}
With these one can give the so-called Iwasawa decomposition $G=H_0 N_{-}H_1$ or $G=H_1 N_{+}H_0$ and the Cartan decomposition $G=H_0 H_{1,+}H_0$. 

The Lie algebra of $\SL(2,\R)$ consists of the traceless $2\times 2$ real matrices, i.e.
\be
\mathfrak{sl}(2,\R)=\{g\in\mathrm{Mat}(2,\R):\tr{(g)}=0\}
\ee
with the commutator acting as the Lie bracket. A basis of the three dimensional vector space $\mathfrak{sl}(2,\R)$ shall be given by $\{h,x,y\}$. The structure of the Lie algebra is then encoded by the commutator relations
\be
[h,x]=2x,~[h,y]=-2y~\text{and}~[x,y]=h.
\ee
In its fundamental representation the generators can be represented by
\be
h=\begin{pmatrix}
1 & 0 \\
0 & -1
\end{pmatrix},~
x=\begin{pmatrix}
0 & 1 \\
0 & 0
\end{pmatrix},~\text{and}~
y=\begin{pmatrix}
0 & 0 \\
1 & 0
\end{pmatrix}.
\ee

$\mathfrak{sl}(2,\R)$ is a simple, particularly a semi-simple Lie algebra. Remarkably, it has two non-conjugated Cartan subalgebras, generated by $x-y$ and $h$~\cite{Lang}. Thus, $\SL(2,\R)$ has two Cartan subgroups, given by the compact group $H_0$, \eqref{eq:H0}, and the noncompact group $H_1$, \eqref{eq:H1}.
(This is to be contrasted to the case of $\mathrm{SL}(2,\mathbb{C})$ which has only one Cartan subgroup.) 

Elements which can be conjugated to a Cartan subgroup are called regular. They form a set which decomposes into the conjugacy classes, specifically
\begin{itemize}
\item[(i)] the \textit{elliptic} classes
\be
G_0=\bigcup_{0<\theta<\pi}\{gu_{\theta}g^{-1}:~g\in G\}
\ee
and the
\item[(ii)] \textit{hyperbolic} classes
\be
G_{\pm}=\bigcup_{t>0}\{g(\pm a_t)g^{-1}:~g\in G\}.
\ee
\end{itemize}

The Haar measure on $\SL(2,\R)$ can then be disintegrated into invariant measures on these classes. Together with the Weyl integration formula, the averaging of a $C_0^{\infty}$-function $f$ over $G$ leads to
\begin{multline}
\int_G \d g f(g)=\\\alpha_0\int_0^{\pi}\d \theta\sin^2\theta f_0(\theta)+\alpha_1\int_0^{\infty}\d t\sinh^2t f_1(t),
\end{multline}
with $f_1(t)\equiv f_1(\pm a_t)$ and $(\alpha_0=1,\alpha_1=1)$, see e.g.~\cite{FreidelLivine}. The functions $f_0$ and $f_1$ denote the averaging of $f$ over the corresponding elliptic and hyperbolic conjugacy classes, that is
\be\label{averagefunctiontheta}
f_0(\theta)=\int_{G/H_0}\d g f\left(g u_{\theta}g^{-1}\right)
\ee
and
\be\label{averagefunctiont}
f_1(t)=\int_{G/H_1}\d g f\left(g (\pm a_{t})g^{-1}\right).
\ee

However, using group averaging arguments, the only way to consistently define an $\text{Ad}(G)$-invariant function $f$ through averaging, is given by the two choices $(\alpha_0=1,\alpha_1=0)$ or $(\alpha_0=0,\alpha_1=1)$. For $L^2$-functions, this amounts to defining two Hilbert spaces $\mathcal{H}_0$ for functions with support on $G_0$ with $(\alpha_0=1,\alpha_1=0)$ and $\mathcal{H}_1$ for functions with support on $G_{\pm}$ and $(\alpha_0=0,\alpha_1=1)$. For a detailed discussion of this point, we refer to Ref.~\cite{FreidelLivine}. Notice that these two sectors cannot be mapped into one another. This can be interpreted as a superselection rule~\cite{Marolf}. In the following subsection we discuss the Fourier decomposition for functions on $G_0$ and $G_\pm$ of the type $f_0(\theta)$ and $f_1(t)$.

\subsection{Harmonic analysis on $\SL(2,\R)$}\label{SLRstuffharmonicanalysis}
 
Here we collect some facts regarding the harmonic analysis on $\SL(2,\R)$ to supplement the main body of this article focusing on the characters and the Plancherel formula. We closely follow Refs.~\cite{SL2Roriginal,Lang,Knapp,Gurarie,GelfandGraevShapiro,SL2Rsonst1,SL2Rsonst2}.

\subsubsection{Characters of $\SL(2,\R)$}\label{characterssl2r}

All unitary irreducible representations of $\SL(2,\R)$ are exhausted by the three series: principal, complementary and discrete. In the following, we give the characters of these and refer to Refs.~\cite{Knapp,Gurarie,SL2Rsonst1,SL2Rsonst2} for their derivation from the respective representations.

$\textbf{1.)}$ The characters of the principal series representation labeled by $s\in\R^+$ are
\begin{multline}
    \chi_{s}^{\pm}(g)=
    \begin{cases}
      \frac{\cos(st)}{|\sinh t|}\epsilon_{\pm}(\lambda_{g}),~\text{for}~
      g~\text{hyperbolic},  \\
      ~~~~~~~~0~~~~~~~~,~\text{for}~
      g~\text{elliptic},
    \end{cases}
  \end{multline}
where $\epsilon_{+}(\lambda_g)=1$ and $\epsilon_{-}(\lambda_g)=\text{sgn}(\lambda_g)$ depending on the eigenvalues $\lambda_g$, \eqref{eq:lambda_g}.

$\textbf{2.)}$ The characters of the complementary series $\chi_{\rho}(g)$ take the same form, only that for these $is$ is replaced by $\rho\in[-1,1]$. Importantly, the complementary series does not contribute to the Plancherel formula (for distributions or $L^2$-functions on $G$)~\cite{Lang,Knapp,Gurarie,GelfandGraevShapiro}.

$\textbf{3.)}$ The characters of the discrete series are
\be
    \chi_{n}^{\pm}(g)=
    \begin{cases}
      \frac{e^{- n |t|}}{2|\sinh t|}
      \epsilon_{\pm}(\lambda_g),~\text{for}~
      g~\text{hyperbolic}, \\
      ~~~~\mp\frac{e^{\pm i n \theta}}{2i\sin\theta}~~~~,~\text{for}~
      g~\text{elliptic},
    \end{cases}
\ee
labeled by $n = 1,2,...$.

The characters are eigenfunctions of the Laplacian with spectrum~\cite{Gurarie}
\be
\frac{1+s^2}{4}  \text{ for }  \chi_{s}^{\pm} \quad\text{and}\quad
\frac{1-n^2}{4}  \text{ for }  \chi_{n}^{\pm} \,. 
\ee
The individual parts of the Laplacian act on averaged functions $f_0(\theta)$ and $f_1(t)$ in the standard way.

\subsubsection{Plancherel formula for $\SL(2,\R)$}\label{SL2RPlancherel}

In view of Appendix~\ref{harmonicanalysisgeneral}, in the case of $\SL(2,\R)$ the inversion formula~\cite{SL2Roriginal,Knapp,Gurarie,SL2Rsonst1,SL2Rsonst2} reads
\ba\label{plancherelinversionsl2r}
f(g)&=&\sum_{n=1}^\infty \frac{n}{4\pi}\left(\tr\left(\hat{f}^{+}(n)\pi_{g}^{n,{+}}\right)+\tr\left(\hat{f}^{-}(n)\pi_{g}^{n,{-}}\right)\right)\nonumber\\
&&+\int_0^\infty\frac{\d s}{4\pi}\frac{s}{2}\tanh\left(\frac{\pi s}{2}\right)\ \tr\left(\hat{f}^{+}\left(s\right)\pi_{g}^{s,{+}}\right)\nonumber\\
&&+\int_0^\infty\frac{\d s}{4\pi}\frac{s}{2}\coth\left(\frac{\pi s}{2}\right)\ \tr\left(\hat{f}^{-}\left(s\right)\pi_{g}^{s,{-}}\right)
\ea
with the Fourier coefficients given by
\be
\hat{f}^{\pm}(s)=\int_G\d g f(g)\pi_{g^{-1}}^{s,{\pm}}
\ee
and
\be
\hat{f}^{\pm}(n)=\int_G\d g f(g)\pi_{g^{-1}}^{n,{\pm}}.
\ee
where $s_\pm$ and $n_\pm$ label the positive and negative branches of the principal and discrete series respectively.
The expression of the inversion formula is due to Harish-Chandra, building on foundational work of Bargmann~\cite{SL2Roriginal,Lang,Knapp}. The first term stems from the discrete series and encapsulates both the contributions coming from the compact and noncompact directions. The second and third terms stem from the continuous series contribution originating from the noncompact directions. 
In particular, this decomposition can be applied to the $\delta$-distribution on $G$~\cite{GelfandGraevShapiro,SL2Rsonst2}, which is simply
\ba\label{deltadistributionsl2r}
\delta(g)&=&\delta_0(\theta)+\delta_1(t)\\
&=&\sum_{n=1}^\infty
\frac{n}{4\pi}\left(\chi_{n}^{+}(\theta)+\chi_{n}^{-}(\theta)\nonumber
+ \chi_{n}^{+}(t)+\chi_{n}^{-}(t)\right)\\
&&+\int_0^\infty\frac{\d s}{4\pi}\frac{s}{2}\left(\tanh\frac{\pi s}{2}\chi_{s}^{+}(t) + 
\coth\frac{\pi s}{2}\chi_{s}^{-}(t)\right)\nonumber
\ea
with $\delta_0(\theta)\equiv\delta_0(u_{\theta})$, $\delta_1(t)\equiv\delta_1(\pm a_{t})$ and the characters are taken as in Appendix~\ref{characterssl2r}. One observes the structural similarities with the case of $\SU(2)$ where the $\delta$-distribution is expanded in terms of characters, see Appendix~\ref{harmonicanalysissu2}.

For functions $f_0(\theta)$ and $f_1(t)$ as given in Appendix~\ref{SLRstuff}, we can use a similar decomposition as Eq.~(\ref{deltadistributionsl2r}), namely
\be
f_0(\theta) = \sum_{n=1}^\infty
\frac{n}{4\pi}\left(\hat{f}^{+}(n)\chi_{n}^{+}(\theta)+\hat{f}^{-}(n)\chi_{n}^{-}(\theta)\right)
\ee
and
\ba
f_1(t) &=&\sum_{n=1}^\infty 
\frac{n}{4\pi}\left(\hat{f}^{+}(n)\chi_{n}^{+}(t)+\hat{f}^{-}(n)\chi_{n}^{-}(t)\right) \nonumber\\
&&+ \int_0^\infty\frac{\d s}{4\pi}\frac{s}{2}\tanh\frac{\pi s}{2}\hat{f}^{+}(s)\chi_{s}^{+}(t) \nonumber\\
&&+ \int_0^\infty\frac{\d s}{4\pi}\frac{s}{2}\coth\frac{\pi s}{2}\hat{f}^{-}(s)\chi_{s}^{-}(t) 
\ea
with Fourier coefficients $\hat{f}^{\pm}(n)$ and $\hat{f}^{\pm}(s)$ for the respective series.

\end{document}